\documentclass[12pt]{book}
\begin{document}

\noindent
{\bf Response to Technical Comments regarding the paper "Direct Detection
of Galactic Halo Dark Matter,'' by B. R. Oppenheimer, N. C. Hambly, 
A. P. Digby, S. T. Hodgkin, D. Saumon, {\it Science} 292, 698 (2001).}

\bigskip
Gibson and Flynn cite several arguments to support their claim that
our study ({\it 1}) greatly overestimated the space density of halo white
dwarfs.  The first stems from a recomputation of our $1 / V_{\rm max}$
calculation.  Their new value actually lies within the uncertainty in
our estimate, however, so it is not clear to begin with that there is
any discrepancy.  Regardless of how the numbers were calculated, the
uncertainties in the space densities are larger than the differences,
and it is not meaningful to discuss them at the accuracies that Gibson
and Flynn demand.  Furthermore, Gibson and Flynn themselves do not
uniformly apply their own criteria for what constitutes an "important"
level of difference.  They clearly view as significant the change
wrought by their initial recalculation, which implies a 14\% decrease
in the calculated white dwarf density, and the change wrought by
applying a white dwarf mass estimate drawn from globular clusters,
which implies an additional 17\% decrease in the density (paragraph 3
of their comment).  Yet they disregard as having "little effect" a
change of 11\% in the opposite direction due to different assumptions
for limiting apparent magnitude (paragraph 4 of their comment).
Finally, the fact that we rounded our final number clearly indicates
that we do not believe it is meaningful to distinguish between 1.6\%
and 2\% based on current data.  Deeper surveys are crucial for
assessing the full extent of this population.

Gibson and Flynn also argue that our study used an average white dwarf
mass, 0.6 M$_\odot$, that was too high by 20\%.  We do not regard this
argument as valid.  Although the average white dwarf mass in globular
clusters is indeed nearer to 0.5 M$_\odot$ than to 0.6 M$_\odot$,
other research suggests that if our findings are correct and the
population of stars represented by the white dwarfs in our study did
emerge from a nonstandard initial mass function, the average white
dwarf mass may actually be much higher.  Chabrier ({\it 2}), for
example, found an average white dwarf mass of 0.7 to 0.8 M$_\odot$.

Gibson and Flynn claim that our results are inconsistent with the LHS
catalog, pointing out that no halo white dwarfs are seen in the
northern portion of the LHS.  Because of the similarities between our
survey and the LHS, they maintain, there can be no halo white dwarfs
in our survey.  That assertion is incorrect, in part because there are
several crucial differences between our survey and LHS and in part
because there clearly are halo white dwarfs in the LHS.  We discuss
these issues in greater detail below, in our response to the Graff
comment.

Finally, Gibson and Flynn conclude that only 0.5\% of the dark matter
is explained by white dwarfs.  The argument through which they reach
that number, however, is not logical.  They take the dimmest star,
WD0351$-$564, and place the entire space density of all of the white
dwarfs found in our survey into the bin of the luminosity function
corresponding to that faintest star---ignoring the fact that it only
accounts for 7\% of the space density.  Then they argue that our survey
did not find enough stars in this faintest luminosity bin. This
circular argument is used to claim that we have overestimated the
space density.  Further, even if one chooses to accept that flawed
argument, one must still contend with the fact that the 0.5\% figure is
only a lower limit.  The complex detection limits (in proper motion
and magnitude) and the incomplete sampling of the velocity parameter
space suggest that a substantially larger population exists and that
we have only found the tip of the iceberg.

Graff's arguments likewise do not withstand close scrutiny.  He begins
with the premise, also proposed by Gibson and Flynn, that our survey
and the LHS catalog are very similar.  As we have already noted, that
premise is incorrect; our survey and the LHS are not directly
comparable.  Our survey reached more than a magnitude deeper and found
many objects to which the LHS was not sensitive.  A clear indication
that we have not simply repeated the LHS work is contained in our
reduced proper motion diagram [figure 1 in ({\it 1})]: The LHS
catalog, as Gibson and Flynn mention, has relatively few objects with
reduced proper motion $H_{\rm R} > 22$; much of our sample, by
contrast, is drawn from the stars with $H_{\rm R} > 22$.  The Liebert
{\it et al.} ({\it 3}) sample of LHS white dwarfs, on which Graff
particularly focuses, likewise included stars with a smaller range of
proper motions than we allowed in our survey.  It is more logical to
interpret these differences as indicating that our survey is deeper
and more complete than to assume that the previous studies of white
dwarfs in the LHS are complete and we have made a mistake.

In addition, one of the coolest white dwarfs known before our survey,
LHS 3250, was first cataloged in the LHS and is notably absent from
the previous studies of the white dwarf content of LHS.  This star,
and other cool white dwarfs that were previously cataloged, had never
been studied spectroscopically before our study.  Rather, the
assumption was that their colors suggested that they were main
sequence stars, not blue, cool white dwarfs with collision-induced
absorption ({\it 4}).  Indeed, to this day, there remain many objects
in the LHS that have not been measured spectroscopically.

These issues render comparison of our survey with the LHS, or the
Liebert {\it et al.} sample of that survey, much more complicated than
simple, back-of-the-envelope calculations based on the tables in
Liebert {\it et al.}, as proposed by Graff.  In reality, a complete Monte
Carlo simulation of the populations and the survey is necessary to
assess the sensitivity of our survey in detail.  Let us assume for the
moment, however, that the Liebert {\it et al.} survey and our survey are
indeed directly comparable, as Graff suggests.  He claims that the
stars we classify as halo stars would simply have been dubbed thick
disk stars by Liebert {\it et al.}  We point out, though, that the Liebert
{\it et al.} sample contained stars that unambiguously are members of the
galactic halo, along with others that may indeed be members of the
halo.  For example, the star LHS 542, which is in their survey, was
clearly shown to be a halo member by Ibata {\it et al.} ({\it 5}).  The
Liebert {\it et al.} ({\it 3}) study is also missing stars: LHS 3250, which
may or may not be a halo member, is conspicuously absent from the
Liebert {\it et al.} study, partly because it was shown to be a white dwarf
only recently ({\it 4}), and Liebert {\it et al.} took known white dwarfs
from LHS without a complete search through the LHS catalog for white
dwarfs that had not been observed in detail.  The halo star WD
0346+246 ({\it 6}, {\it 7}) is missing from both Liebert {\it et al.} ({\it
3}) and the LHS, even though it was within the photometric and proper
motion detection limits of the LHS---which suggests that LHS is not
complete at the R = 18.5 level for white dwarfs.  The latter two stars
both have peculiar spectral energy distributions.

In short, the LHS survey certainly does contain dark halo white dwarfs
that have not previously been identified as such.  Indeed, 11 of the
38 stars that we listed [table 1 in ({\it 1})] were in the LHS or LP
catalog.  However, the LHS catalog contains only a small number of
halo white dwarfs, and it is certainly not complete at the detection
levels necessary to reveal a convincing fraction of the halo white
dwarf population.

Graff continues by claiming that because the local density of halo
main sequence stars is 600 times smaller than the density of disk main
sequence stars ({\it 8}), the same should be true for the white dwarfs.
That claim has no basis in our current understanding of these two
different populations of stars.  First of all, if the halo is composed
of substantially older stars that formed roughly coevally, as is
generally believed, one would expect substantially different ratios of
white dwarfs and main sequence stars in the halo and disk.  The disk
is believed to be a population of stars that have been continuously
forming since the disk formed.  The comparison is thus moot, and the
assertion that 1 in 30 of the stars in our sample may be halo white
dwarfs is rendered incorrect.  The first examples of halo white dwarfs
were discovered convincingly only in the past few years, and the
construction of relative numbers of these stars is impossible if one
disregards the results that we published in ({\it 1}).

According to Graff, because our survey is proper-motion limited, we
necessarily have included more of the disk stars than we thought.
Effectively, he claims, our 2-$\sigma$ exclusion is relegated to a
1-$\sigma$ exclusion of the disk stars.  That statement clearly does
not hold in all cases and, most important, in this case.  Halo and
even thick-disk stars should have average proper motions higher than
those of the disk; their kinematics are necessarily different from
those of the Sun.  Furthermore, the 94 km s$^{-1}$ number---which is
actually centered at the point $(V,U) = (-35,0)$, not $(V,U) = (0,0)$
as Graff seems to have assumed---comes from the survey by Chiba and
Beers ({\it 9}), which examined the velocity distributions of stars
that were not kinematically selected.  Therefore, there is no question
that 94 km s$^{-1}$ is a 2-$\sigma$ value.

To respond to Graff's final point, we have not yet assessed the
sensitivity of the survey in ({\it 1}) as a function of proper motion
with any accuracy.  [We did point out in ({\it 1}) that there was a
less than 10\% chance that we would find any stars with 3 arcseconds
of motion per year or greater.]  To assess that sensitivity---and, more
important the sensitivity of our survey in the $VU$ parameter space that
we plotted [figure 3 in ({\it 1})]---will require detailed modeling of
the survey and the various galactic populations.

\medskip\noindent
{\bf B. R. Oppenheimer}

Astronomy Department,
University of California, Berkeley,

Berkeley, CA 94720-3411, USA,
E-mail:  bro@astron.berkeley.edu

\medskip\noindent
{\bf N. C. Hambly, A. P. Digby}

Institute for Astronomy,
University of Edinburgh,
Royal Observatory,

Blackford Hill,
Edinburgh, EH9 3HJ, UK

\medskip\noindent
{\bf S. T. Hodgkin}

Institute of Astronomy,
Cambridge University,
Madingley Road,

Cambridge, CB3 0HA, UK

\medskip\noindent
{\bf D. Saumon}

Department of Physics and Astronomy,
Vanderbilt University,

Nashville, TN 37235, USA

\bigskip

\noindent
{\bf References and Notes}

\noindent
1.  B. R. Oppenheimer, N. C. Hambly, A. P. Digby, S. T. Hodgkin, D. Saumon, {\it Science} 292, 698 (2001).

\noindent
2.  G. Chabrier, {\it Astrophys. J. Lett.} 513, L103 (1999).

\noindent
3.  J. Liebert, C. Dahn, D. G. Monet, {\it Astrophys. J.} 332, 891 (1988).

\noindent
4.   H. C. Harris {\it et al.}, {\it Astrophys. J.} 524, 1000 (1999). 

\noindent
5.   R. A. Ibata {\it et al.}, {\it Astrophys. J. Lett.} 532, L41 (2000).

\noindent
6.  N. C. Hambly {\it et al.}, {\it Astrophys. J. Lett.} 489, L157 (1997).

\noindent
7.  S. T. Hodgkin {\it et al.}, {\it Nature} 403, 57 (2000).

\noindent
8.  This point is actually a major source of contention among experts
    in this field and the number varies from Graff's 600 to well below
    250 [({\it 9}); also ({\it 10}) and references therein].

\noindent
9.  M. Chiba and T. C. Beers, {\it Astron. J.} 119, 2843 (2000).

\noindent
10.  Gould, C. Flynn, J. N. Bahcall, {\it Astrophys. J.} 503, 798 (1998).

\bigskip\noindent
4 May 2001; accepted 18 May 2001

\end{document}